\documentclass[12pt]{iopart}

\usepackage{harvard}

\usepackage{graphicx}

\begin{document}

\title[The Galactic Stellar Disc]{The Galactic Stellar
  Disc\footnote{This paper includes data gathered with the 6.5 meter
    Magellan Telescopes located at Las Campanas Observatory, Chile. 
Also based on observations collected at the Nordic Optical Telescope 
on La Palma, Spain, and at the European Southern Observatories on 
La Silla and Paranal, Chile, Proposals \#65.L-0019(B), 67.B-0108(B), 
69.B-0277}}

\author{S Feltzing$^1$, T Bensby$^2$}

\address{$^1$ Lund Observatory, Box 43, SE-221~00 Lund, Sweden}
\address{$^2$ European Southern Observatory, Alonso de Cordova 3107,
  Vitacura, Casilla 19001, Santiago, Chile}

\ead{sofia@astro.lu.se}

\begin{abstract}
  The study of the Milky Way stellar discs in the context of galaxy
  formation is discussed. In particular we explore the properties of
  the Milky Way disc using a new sample of about 550 dwarf stars for
  which we have recently obtained elemental abundances and ages based
  on high resolution spectroscopy. For all the stars we also have full
  kinematic information as well as information about their stellar
  orbits. We confirm results from previous studies that the thin and
  the thick disc have distinct abundance patterns.  But we also
  explore a larger range of orbital parameters than what has been
  possible in our previous studies. Several new results are
  presented. We find that stars that reaches high above the galactic
  plane and have eccentric orbits show remarkably tight abundance
  trends. This implies that these stars formed out of well mixed gas
  that had been homogenized over large volumes. We find some evidence
  that point to that the event that most likely caused the heating of
  this stellar population happened a few billion years ago.  Through a
  simple, kinematic exploration of stars with super-solar [Fe/H] we
  show that the solar neighbourhood contains metal-rich, high velocity
  stars that very likely are associated with the thick disc.
  Additionally, the HR1614 moving group and the Hercules and Arcturus
  stellar streams are discussed and it is concluded that, probably, a
  large fraction of the so far identified groups and streams in the
  disc are the result of evolution and interactions within the stellar
  disc rather than being dissolved stellar clusters or engulfed dwarf
  galaxies.
\end{abstract}

\pacs{97.20.Jg, 98.10.+z, 98.20.-d, 98.35.Ac, 98.35.Bd, 98.35.Hj, 98.35.Pr}


\section{Introduction}

The formation and evolution of galaxies is a key topic in contemporary
astrophysics and a major driver for large observational facilities,
such as the E-ELT, as well as extra-galactic surveys, one example
being the Sloan Digital Sky Survey (SDSS). However, also the study of
the very local universe in the form of the Milky Way and the Local
Group are important in the context of galaxy formation.  Indeed,
$\Lambda$CDM, the currently most successful theory for the formation
of large scale structure in the universe \cite{2006Natur.440.1137S},
meets with some important challenges when the smallest scales are
being studied \cite{2002ARA&A..40..487F,2008arXiv0805.1244D}. These
tests include questions relating for example to the number of dwarf
galaxies and the formation and evolution of disc systems in spiral
galaxies.

The formation and survival of disc galaxies within $\Lambda$CDM has
recently been the focus of several investigations
\cite{2008arXiv0803.2714R,2008arXiv0806.2861H,2008arXiv0806.2872D}.
Given the hierarchical nature of the build up of galaxies within
$\Lambda$CDM it seems natural that the formation of old discs might be
a problem as what today is a giant galaxy must have suffered a
bombardment of merging blocks over its lifetime. However, recent
studies imply that the effects might not be as severe as initially
thought. Two major issues have recently been
discussed. \citeasnoun{2008arXiv0806.2872D} investigated the merger
trees for Milky Way like halos in the Millennium Simulation\footnote{A
  description of the simulations as well as access to the publicly
  available database from the simulation can be found at
  http://www.mpa-garching.mpg.de/millennium/} and found that a
majority of these indeed had very calm merger histories -- allowing a
build up of discs that would not be destroyed by major
mergers. Although the large scale structure simulations can give us
the number of mergers and interactions that a galaxy experiences
throughout the age of the universe the effect that a merger or close
interaction can have on a baryonic disc needs to be modeled separately
in order to fully understand the damage in e.g. the form of heating of
a pre-existing thin stellar disc to a thick disc.
\citeasnoun{2008arXiv0806.2861H} studied the heating effect of mergers
on a large disc and found that the merger rate could be higher than
previously thought without destroying the disc. The major reason for
the increased number of allowed, non-destructive mergers is the fact
that previous studies, that showed that smaller numbers already
destroyed the disc, have not taken the full range of possible orbital
parameters for merging satellites into account. By allowing for all
 types of orbits as well as using realistic sizes for the halos of
the merging galaxies (as realized in cosmological simulations)
\citeasnoun{2008arXiv0806.2861H} found that the Milky Way could have
had about 5 to 10 1:10 mass-ratio mergers since $z\sim2$, which is in
agreement with cosmological simulations.

We are thus in a situation where the simulations indicate on the one
hand that Milky Way like galaxies must live in unusual places in the
Universe in order to survive, on the other hand other types of
modeling indicate that a Milky Way can survive a ``typical''
cosmological environment. The study of both local galaxies as well as
galaxies far away can thus provide valuable data to constrain future
modeling.

There is mounting observational evidence that disc galaxies exist at
high redshifts. A recent result is the study of velocity maps for 11
galaxies at $z\sim 2$ by \citeasnoun{2008arXiv0802.0879S} using the
integral field spectroscopy instrument SINFONI on VLT. Their
observations show that more than half of the galaxies in their sample
have not suffered any recent major merger and the galaxies show ordered disc
motions in the H$\alpha$ gas. The dynamical masses of these galaxies
are large, $10^{10-11}$M$_{\odot}$. Several other studies have also
shown the existence of both old stars as well as solar metallicities
at similar redshifts. Two examples are given by
\citeasnoun{2004ApJ...605...37S} and \citeasnoun{2004ApJ...612..108S}.

In the more nearby universe thick discs have been shown to be near
ubiquitous in edge-on galaxies \cite[and references
therein]{2006AJ....131..226Y}. In a recent study
\citeasnoun{2008arXiv0805.4197Y}, using Lick indexes, study the ages
and metallicities of the stellar populations in a sample of edge-on
galaxies with thick discs and found that the extra-planar regions in
these galaxies are systematically older and less metal-rich than the
stellar population in the mid-plane.

The most nearby disc system we can study is that in the Milky Way
galaxy.  The Milky Way has two stellar discs -- one thick and one
thin. These two discs differ in their properties. In particular, the
thick disc is older and less metal-rich than the thin disc
\cite{2002ARA&A..40..487F}. The thick disc also rotates more slowly
around the centre of the Galaxy than the thin disc. The so called
asymmetric drift is $\sim50$ km\,s$^{-1}$ for the thick disc and
$\sim15$ km\,s$^{-1}$ for the thin disc
\cite{2002ARA&A..40..487F}. The thick disc is generally thought to
have a scale-height about 3 times larger than that of the thin
disc. However, the exact scale-height of the thick disc as well as the
local normalization in the solar neighbourhood, i.e. how many thick
disc stars there are per thin disc star, is still not
well-established. The values determined for the scale height for the
thick disc found in the literature ranges from 600 pc to more than
1600 pc and the local normalization from 12\,\% to about 2\,\%. A
compilation of the various determinations of the thick disc scale
height as well as its local normalization can be found in
\citeasnoun{2008arXiv0807.1665A}.

In the context of galaxy formation and as tests of models of galaxy
formation it is important to establish the properties of a disc system
of the type observed in the Milky Way. Studies of nearby galaxies as
well as galaxies at higher redshifts show evidence for old,
well-ordered gas and stellar discs. Some of these discs are also
thick. The study of the Milky Way discs thus provides the very local
example of such a system.

\section{Kinematics, orbital parameters and elemental abundances}
 \begin{figure}
   \centering
   \resizebox{\hsize}{!}{\includegraphics[]{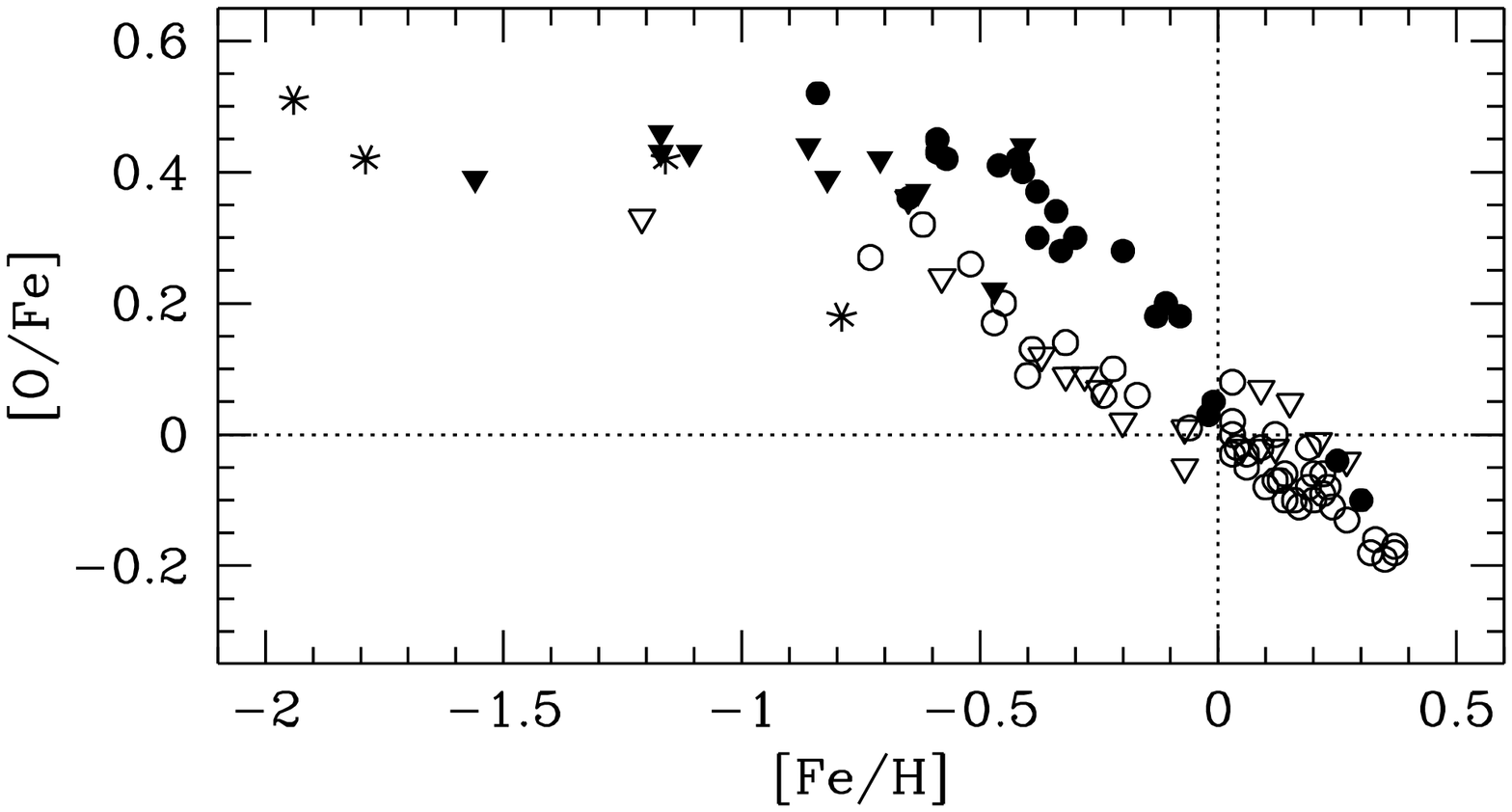}}
   \caption{Oxygen abundances for stars with kinematics typical of the
     thin disc ($\circ$) and of the thick disc ($\bullet$)
     \cite{2004A&A...415..155B} and stars with kinematics typical of
     the thin disc ($\bigtriangledown$), the thick disc (filled
     triangles) and halo ($\ast$) \cite{2002A&A...390..235N}. The data
     from \citeasnoun{2004A&A...415..155B} is based on observations of
     the [OI] line at 630.0 nm with the Coud\'e Echelle Spectrograph
     (CES) on the ESO 3.6m. The observations used the highest
     resolving power of the CES (R$\simeq$215\,000) and the spectra
     have very high signal-to-noise ($\geq 400$). }
   \label{oxygen}
   \end{figure}

One of the first studies to systematically explore the potential of
combining elemental abundances for stars with detailed knowledge about
their ages, kinematics and orbital parameters was done by
\citeasnoun{1993A&A...275..101E}. A major outcome of this work was to
firmly establish that the Milky Way disc, in fact mainly the thin
disc, shows tight and well defined abundance trends for a large range
of elements, such as Si, Ca, and Ni.

In 1997 the Hipparcos catalogue became available
\cite{1997A&A...323L..49P}. Combining this catalogue of parallaxes and
proper motions with radial velocities from the literature it became
possible to trawl large databases for stars with kinematics typical of
the thin and the thick discs. \citeasnoun{2003A&A...410..527B} used
this possibility and constructed two samples of F and G dwarf stars
representative of the thin and the thick discs and showed that these
two stellar samples differ. For example, the stars in the thick disc
are more enhanced in [O/Fe], at a given [Fe/H], than the stars in the
thin disc, \fref{oxygen}. A first, simplistic, interpretation of the
trend found for the thick disc stars is that the thick disc shows
evidence of contribution by SN\,Ia to its chemical evolution as
evidenced by the presence of the ``knee'' in the [O/Fe] vs. [Fe/H]
trend. This trend is typically interpreted such that the constant
[O/Fe] is the result of enrichment of the interstellar gas from
SN\,II.  These supernovae form both oxygen and some iron and when the
star forming gas sees the integrated enrichment from several
supernovae the result is a constant oxygen to iron ratio. Iron is also
produced in SN\,Ia but oxygen is not. 
Hence, when SNIa start to contribute to the enrichment of the 
interstellar medium the oxygen to iron ratio will decrease
in the subsequently formed stars.
The time-scale for
SN\,Ia are not firmly established, however, it is not only the
life-time for a single object that is of interest but more so the
formation rate of these types of systems and then their integrated
effect on the chemical enrichment. The lifetimes for SN\,II are short
and it is generally expected that the SN\,Ia enrichment happens on a
longer time-scale, however, in certain cases it can also happen on
time-scales shorter than the canonical one billion years. For a deeper
discussion of these issues and further references see the recent
lecture notes by \citeasnoun{2008arXiv0804.1492M}.

An obvious limitation in the identification of thin and thick disc
stars through the kinematic probabilities as used in e.g
\citeasnoun{2003A&A...410..527B} is that the interpretation requires
that a star's history is, at least to some extent, retained in its
current kinematic properties.  In this context the study by Fuhrmann
are important \cite{2008MNRAS.384..173F,1998A&A...338..161F}. He
studied a volume limited sample ($d<25$ pc) of dwarf stars in a narrow
temperature range (essentially late F and early G dwarf stars). The
sample selection does not include any consideration of the kinematic
properties of the stars. The stars in this volume limited sample show
two clear, and separate trends for [Mg/Fe] as a function of [Fe/H].
Essentially, there is one set of stars that are enhanced in [Mg/Fe]
and one that is not. The stars that are enhanced in [Mg/Fe] fall on a
plateau in the abundance diagram and the most metal-rich of them have
[Fe/H] of about --0.3 dex. Fuhrmann identifies these stars with the
thick disc (compare the enhanced [O/Fe] in \fref{oxygen} for the,
kinematically identified, thick disc stars). The other trend contains
stars that are younger and with lower [Mg/Fe]. The [Mg/Fe] as a
function of [Fe/H] trend for these stars is comparable to the oxygen
trend for the kinematically selected thin disc stars in \fref{oxygen}.

Most importantly, the study by \citeasnoun{2008MNRAS.384..173F} shows
that a volume limited sample gives the same resulting abundance trends
as found in the kinematically selected samples. Fuhrmann's study thus
provides a validation of the selection techniques developed in the
literature. These techniques use various kinematic criteria that are
turned into membership probabilities that are then used to identify
the thin and thick disc stars. Examples include the methods developed
in the following studies \citeasnoun{2003A&A...406..131G},
\citeasnoun{2003MNRAS.340..304R}, \citeasnoun{2003A&A...410..527B},
\citeasnoun{2004AJ....128.1177V}, \citeasnoun{2005A&A...438..139S},
and \citeasnoun{2006MNRAS.367.1329R}.

\subsection{A new stellar sample}
\label{sample}

   \begin{figure}
   \centering
   \resizebox{\hsize}{!}{\includegraphics[]{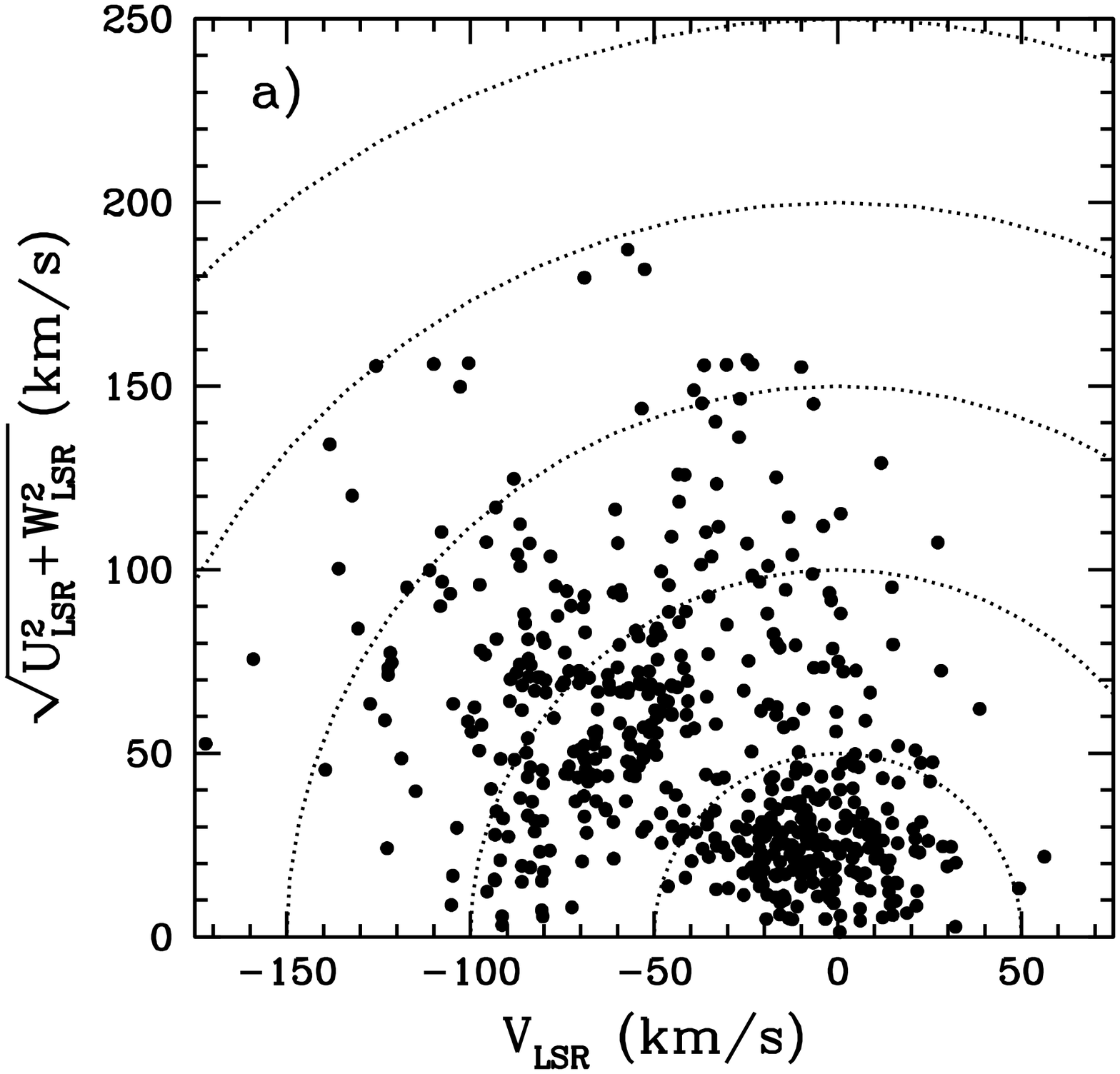}\includegraphics[]{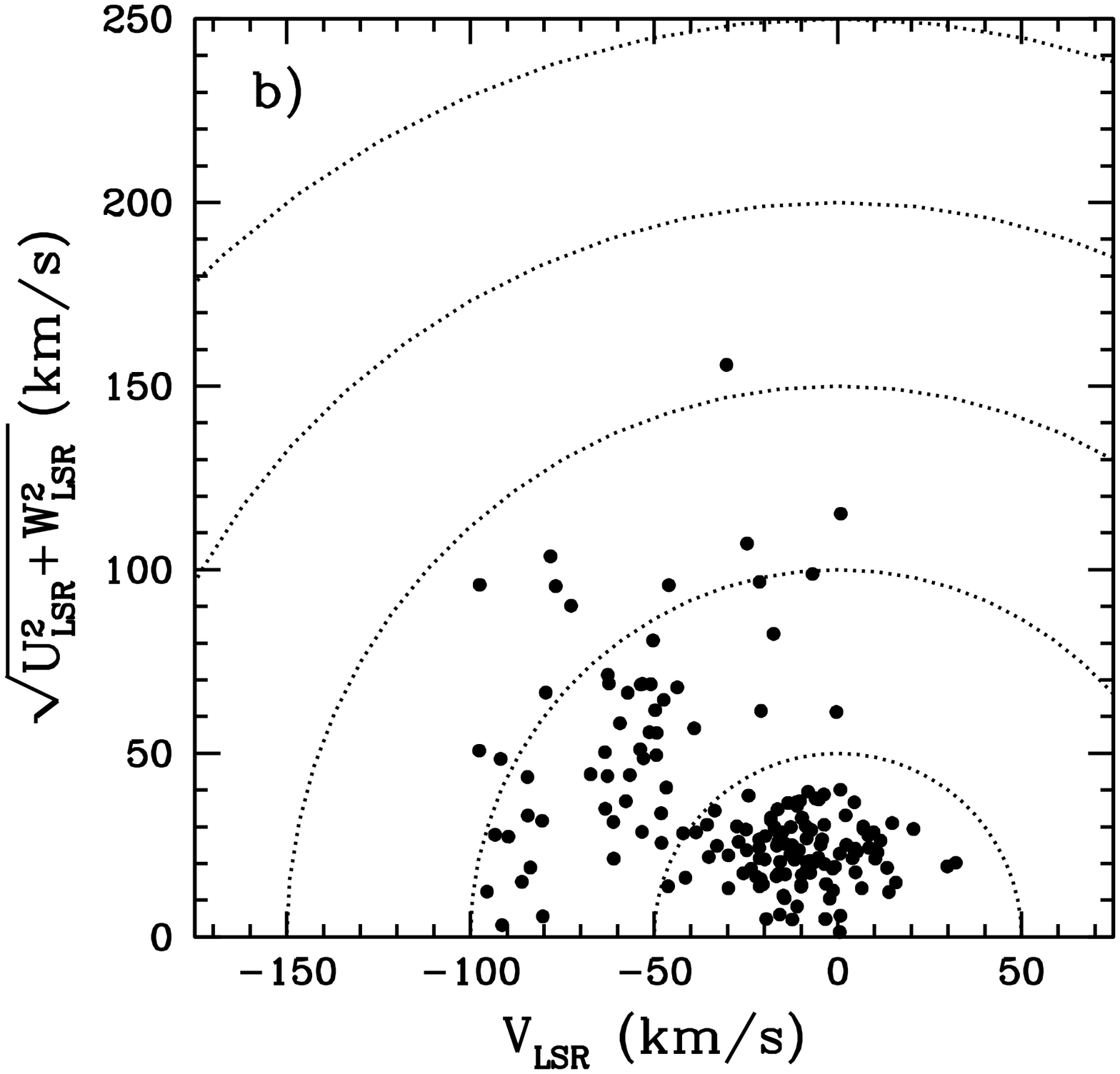}}
   \caption{{\bf a.} Toomre diagram for the whole sub-sample of $\sim$550
     stars. {\bf b.} Toomre diagram only for stars with [Fe/H]$>$0. In
     both panels the concentric circles, shown with dotted lines,
     indicate a constant total velocity in steps of 50
     km\,s$^{-1}$.}
   \label{figone}
   \end{figure}

\label{explore}
  \begin{figure}
   \centering
   \resizebox{\hsize}{!}{\includegraphics[]{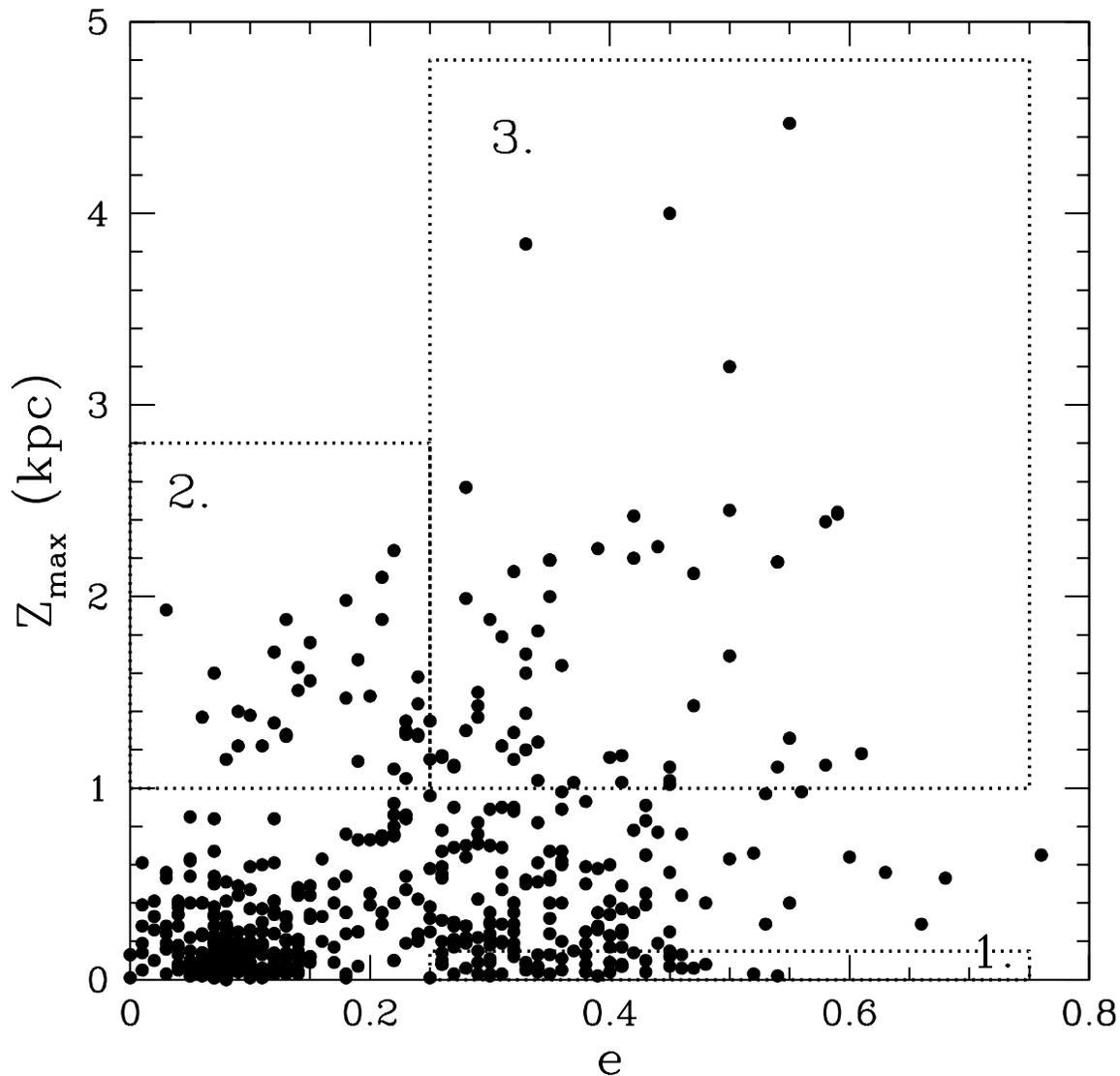}}
   \caption{$Z_{max}$ as a function of $e$ for the stellar sample.
     Three boxes show the outlines of the parameter spaces considered
     in \fref{figthree} and \sref{explore}. The numbers for the boxes
     corresponds to the same numbers for the panels in \fref{figthree}
     and \ref{bathree}.  The orbital parameters are taken from
     \citeasnoun{2004A&A...418..989N}.}
   \label{figze}
   \end{figure}

  \begin{figure}
   \centering
   \resizebox{\hsize}{!}{\includegraphics[]{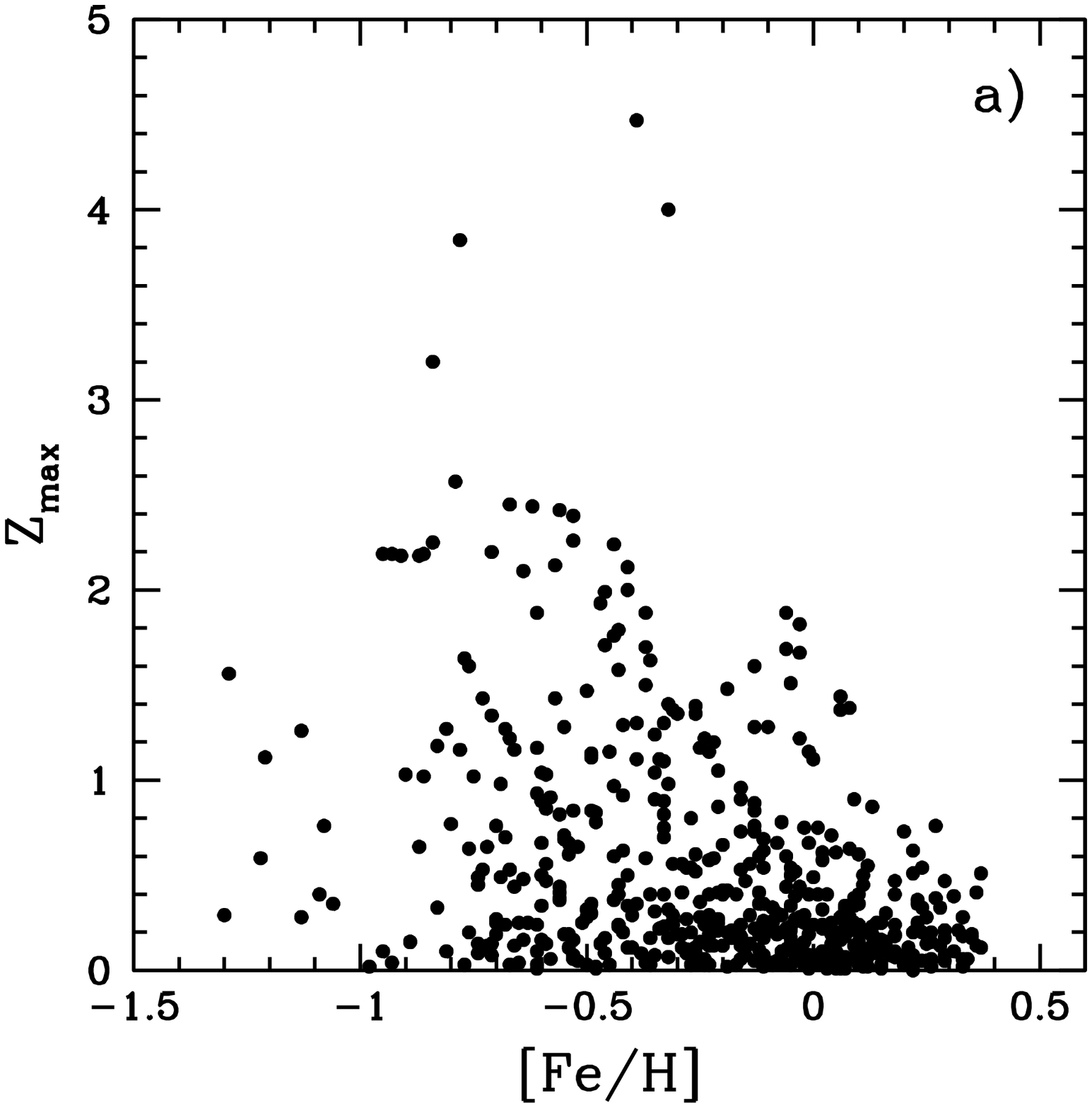}\includegraphics[]{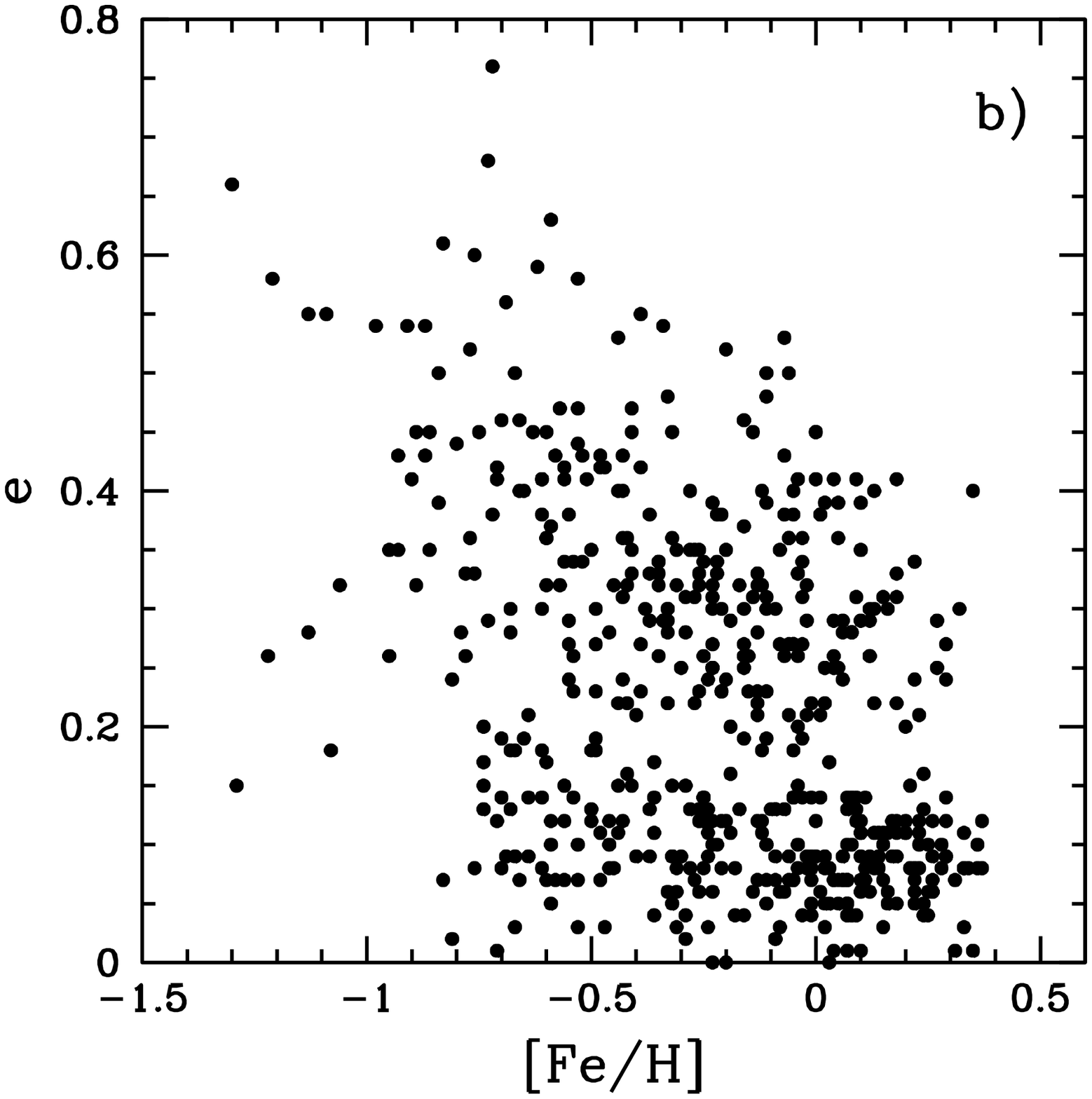}}
   \caption{{\bf a.} $Z_{max}$ as a function of [Fe/H] for our
     sample. {\bf b.} $e$ as a function of [Fe/H] for our sample. The
     apparent gap at $e\sim0.2$ is an artifact resulting from the
     selections of the various sub-samples that make up the final
     sample (see discussion in \sref{sample}). The orbital parameters
     are taken from \citeasnoun{2004A&A...418..989N}.}
   \label{figzefeh}
   \end{figure}

   We have obtained high-resolution, high S/N spectra for about 900
   dwarf stars. The data have been obtained with several spectrographs
   but in general S/N$>$250 and R$\geq$65,000 (apart from the sub-set
   of stars originally observed with FEROS \cite{2003A&A...410..527B}
   which have R=48,000). In this publication we present the kinematic
   properties and some elemental abundances as well as discuss the
   stellar ages for a sub-sample of about 550 F and G dwarf stars. We
   employ the same methods of abundance analysis as outlined in
   \citeasnoun{2003A&A...410..527B}. However, in the present analysis
   we allow for enhancement of $\alpha$-elements in the model
   atmospheres and we use a somewhat more automatic process to
   determine the stellar parameters. The surface gravity and age
   determinations are still based on the Hipparcos parallax
   \cite{1997A&A...323L..49P} but now using the parallaxes from the
   new reduction provided by \citeasnoun{2007hnrr.book.....V}.  For
   the present discussion we use the orbital parameters from
   \citeasnoun{2004A&A...418..989N}. However, given that for some
   stars the new parallaxes and proper motions are sufficiently
   different from those in the original Hipparcos catalogue we will
   re-calculate them. We have already derived new stellar ages based
   on our own effective temperatures and the new parallaxes. For the
   derivation of stellar ages we use the Yonseii-Yale isochrones
   \cite{2002ApJS..143..499K,2004ApJS..155..667D}.

   \Fref{figone} shows the kinematic properties of the $\sim$550
   stars we present here. The full sample is built from several
   sub-samples. Each of these sub-samples have originally been
   selected to study a particular issue, e.g. how metal-rich can the
   thick disc be \cite{2007ApJ...663L..13B}. The result is a somewhat
   uneven distribution in velocity space. This is perhaps more obvious
   in \fref{figze} which shows the maximum height above the galactic
   plane ($Z_{max}$) that a star reaches as a function of the
   ellipticity of the orbit for the star ($e$). One aspect of our
   sample that sets it apart from e.g. the studies by
   \citeasnoun{2006MNRAS.367.1329R} and
   \citeasnoun{2003MNRAS.340..304R} is that we cover a wider range of
   orbital parameters.  For example our sample includes stars on low
   $e$ orbits with low [Fe/H] (chosen to study the metal-weak thin
   disc) as well as stars with super-solar [Fe/H] and high $e$ (chosen
   to study the metal-rich thick disc). Neither of these types of
   stars have been systematically included in previous studies. In
   fact in some studies they are lacking all together and the
   inclusion of them in our study has enabled us to explore a wider
   range of the parameter space (compare \fref{figzefeh}).

\subsection{Exploration of the Milky Way stellar disc}
   \begin{figure}
   \centering
   \resizebox{\hsize}{!}{\includegraphics[angle=-90]{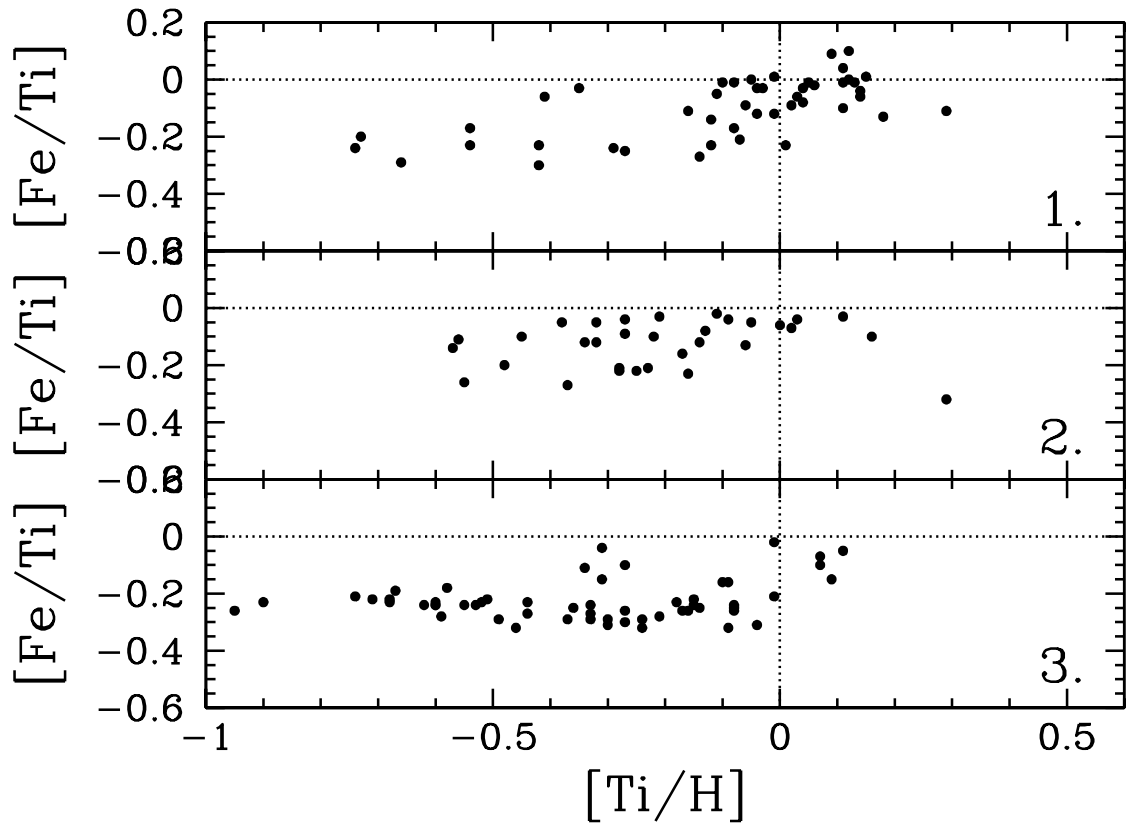}}
   \caption{[Fe/Ti] vs. [Ti/H] for the three samples defined in
     \fref{figze}. The numbering of the panels corresponds to the
     numbering of the boxes in \fref{figze}.}
   \label{figthree}
   \end{figure}

   \begin{figure}
   \centering
   \resizebox{\hsize}{!}{\includegraphics[angle=-90]{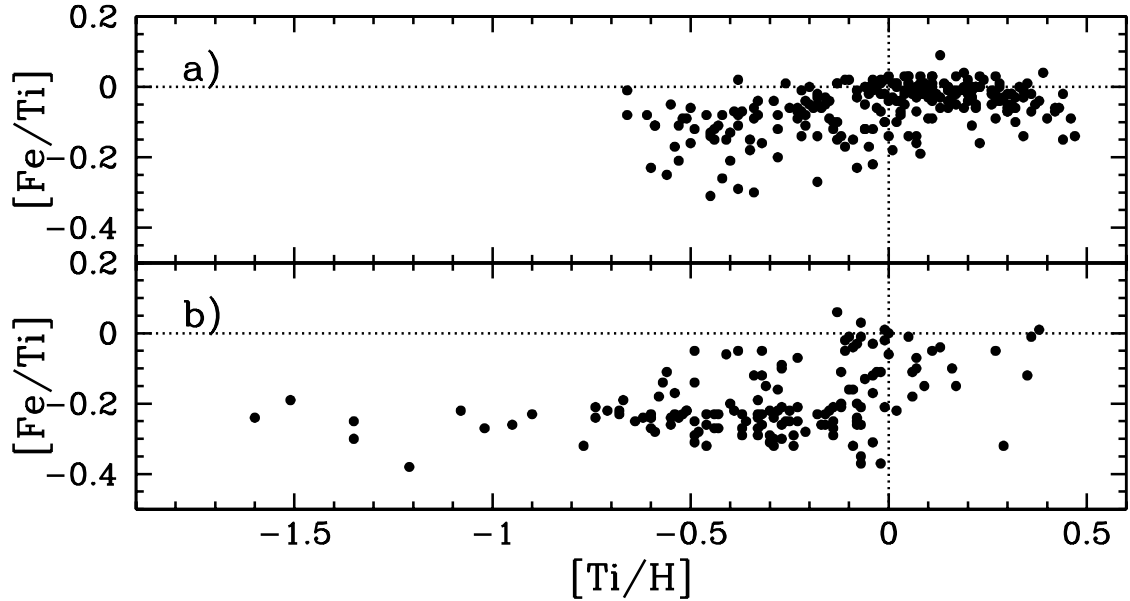}}
   \caption{[Fe/Ti] vs. [Ti/H] for {\bf a.} stars that are ten times
     more likely to be thin than thick disc and {\bf b.} stars that
     are ten times more likely to be thick than thin disc stars.  The
     separation is based on probabilities calculated according to the
     prescriptions in \citeasnoun{2003A&A...410..527B}. }
   \label{figtt}
   \end{figure}

   We start our exploration of the Milky Way by considering the
   $Z_{\rm max}$-$e$ space.  In \fref{figze} three areas of interest
   are indicated. All three boxes contain what could be deemed thick
   disc stars. Box 1 includes the stars that move on very eccentric
   orbits but have low $W_{\rm LSR}$, Box 2 stars that have low
   eccentricities but high $W_{\rm LSR}$, and Box 3 includes
   the stars that have both high $e$ and high $W_{\rm
     LSR}$. \Fref{figthree} then shows the elemental abundance trends
   for the stars that fall in these three boxes. We have chosen to
   show [Fe/Ti] vs [Ti/H] as the separation between e.g. thin and
   thick disc tends to be most obvious for this combination of
   elements but also because Ti is essentially made in SN\,II whilst
   Fe comes from both SN\,II and Ia. Hence we now have a clearer
   tracer of time on the $x$-axes than when using Fe as the reference
element.

The trends for the three boxes differ. Box 1 and 3 are quite similar,
although stars in Box 3 show a more distinct trend. For both samples
we see a flat trend with a constant [Fe/Ti] at about --0.25 dex that
extends from [Ti/H] of --1 to 0 dex. The trend for the stars in Box 3
shows a possible upturn in [Fe/Ti] after [Ti/H]=0. For the stars in
Box 1, however, there is a clear upturn around [Ti/H]=0.  In contrast,
the stars from Box 2 show a fairly flat distribution with a higher
[Fe/Ti] (mean value of roughly --0.1 dex) and a larger spread in
abundance ratios.

The tightness of the trend seen for the stars in Box 3 is quite
remarkable. The star-to-star scatter is less than 0.1 dex over a range
of 1 dex (omitting four stars at [Ti/H] = --0.3). Such a trend can
only be achieved if the stars formed out of a gas that was well mixed
\cite{2004astro.ph.11714G}. Moreover, taking the orbital parameters
for the stars in Box 3 into account this implies that the gas these
stars formed out of had not only seen the average effect of about a
hundred supernovae but that the mixing was efficient over a volume
spanning several kpc.

The trend for the stars in Box 3 and 1 can be compared with the trend
found for the thick disc using the selection criteria in
\citeasnoun{2003A&A...410..527B}. Their selection method is based on
the fact that the thin disc, thick disc, and halo rotate at different
speeds around the centre of the Galaxy and that for each of these
stellar populations we know the velocity dispersions in UVW. Using
this information it is possible to calculate a probability that a star
belongs to one of the three components. These probabilities can then
be compared and very likely thin and thick disc members, respectively,
can be picked.  \Fref{figtt}b shows the thick disc [Fe/Ti] vs. [Ti/H]
abundance trend using this selection technique. Only stars that are
ten times more likely to be thick than thin disc stars are shown.  The
trend found for the thick disc with this selection is more scattered
in appearance but has the same basic features as the much more
well-defined trend found in Box 3. This is, most likely, an indication
of the limitations of the statistical techniques we use to select the
thin and the thick disc stars.  However, also here we see an
essentially flat and rather tight abundance trend spanning over one
dex in [Ti/H]. For $-0.6 < $[Ti/H]$ <-0.2$ there is some likely thin
disc contamination. Disregarding these stars we are again led to the
conclusion that the thick disc stars must have formed out of
well-mixed gas that had seen the integrated effect of many
supernovae.

The elemental abundance trend for stars selected in Box 2,
\fref{figze}, can then be compared with that for stars selected to be
very likely thin disc stars, as shown \fref{figtt}a. Although the
number of stars in Box 2 are few and hence the trend not so well
established as for the thin disc sample it is clear that the two
trends agree extremely well and that both are distinct from that of
the thick disc as well as being distinct from the trends found for the
stars in Box 3 and 1.

\subsubsection{Stellar ages}

  \begin{figure}
   \centering
   \resizebox{\hsize}{!}{\includegraphics{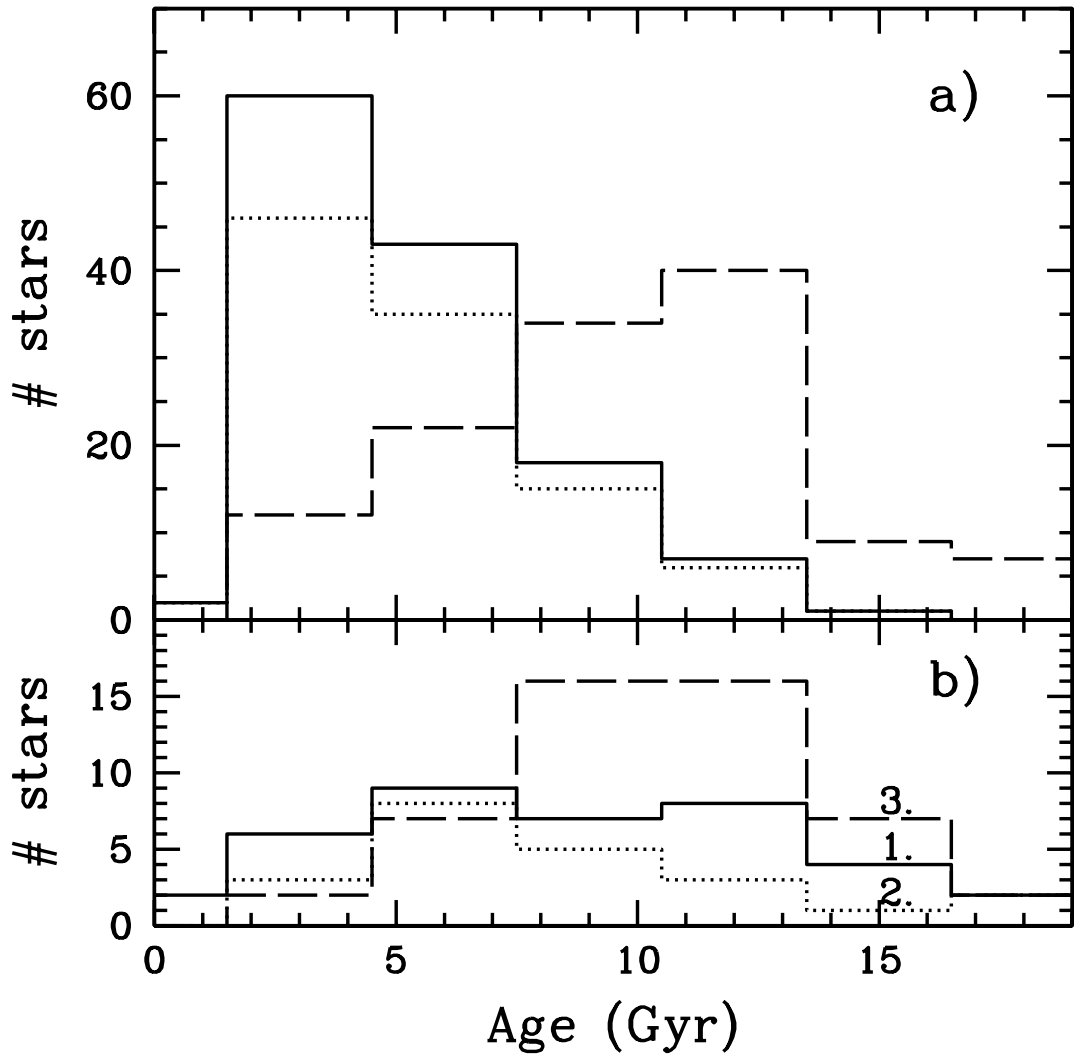}}
   \caption{Distribution of stellar ages for the samples of stars
for which the elemental abundance trends are
     shown in \fref{figtt} and \fref{figthree} and identified in
     \fref{figze}. In these histograms we use the stars with
     ``good'' age determination. For the current sample this means
     that we have only included stars for which the internal, relative
     error in the age determination is less than 30\%. {\bf a.} Age
     distribution for the stars selected to be ten times more likely
     to be thin than thick disc stars is shown with a full line. The
     distribution for stars that are ten times more likely to thick
     than thin disc stars are shown with a dashed line (compare
     \fref{figtt}b). The dotted line shows the distribution for stars
     with $0<e<0.15$ and $Z_{\rm max}<0.5$ kpc (compare
     \fref{figze}). {\bf b.}  This panel shows the age distributions
     for the three stellar samples selected in Box 1 -- 3 in
     \fref{figze} and for which the elemental abundance trends are
     shown in \fref{figthree} and \ref{bathree}. The histograms are labeled with their
     respective box identification as used in those figures.}
   \label{figages}
   \end{figure}

The distributions of the ages for the stars selected in the three
boxes and for the thin and the thick disc are shown in \fref{figages}.
We only include stars for which the age could be determined with a
relative error less than 30\%.  The thin and the thick discs show
overlapping distributions but with distinct mean ages, the
thick disc is on average older than the thin disc.  The thin and the
thick discs are selected according to the same criteria as for the
abundance trends shown in \fref{figtt}. We also show the age
distribution for stars on very circular, planar orbits ($0<e<0.15$ and
$Z_{\rm max}<0.5$ kpc). This could be regarded as a very conservative
thin disc sample. It is reassuring to see that the age distribution
for those stars is essentially the same as for our thin disc sample.

The stars from Box 2 show an age distribution skewed to younger
ages. Box 1 has an essentially flat age profile and Box 3 shows a
distribution similar to that of the thick disc. The number statistics
are low for Box 1 and 2 and any conclusions drawn from this should be
regarded as tentative and needs confirmation with larger samples.  Our
full sample of 900 stars indeed includes many more stars on such
orbits that still needs to be analyzed and should help clarify the
picture.

\subsubsection{Additional evidence for well-mixed gas -- Barium abundances}
\label{barium}

  \begin{figure}
   \centering
   \resizebox{\hsize}{!}{\includegraphics[angle=-90]{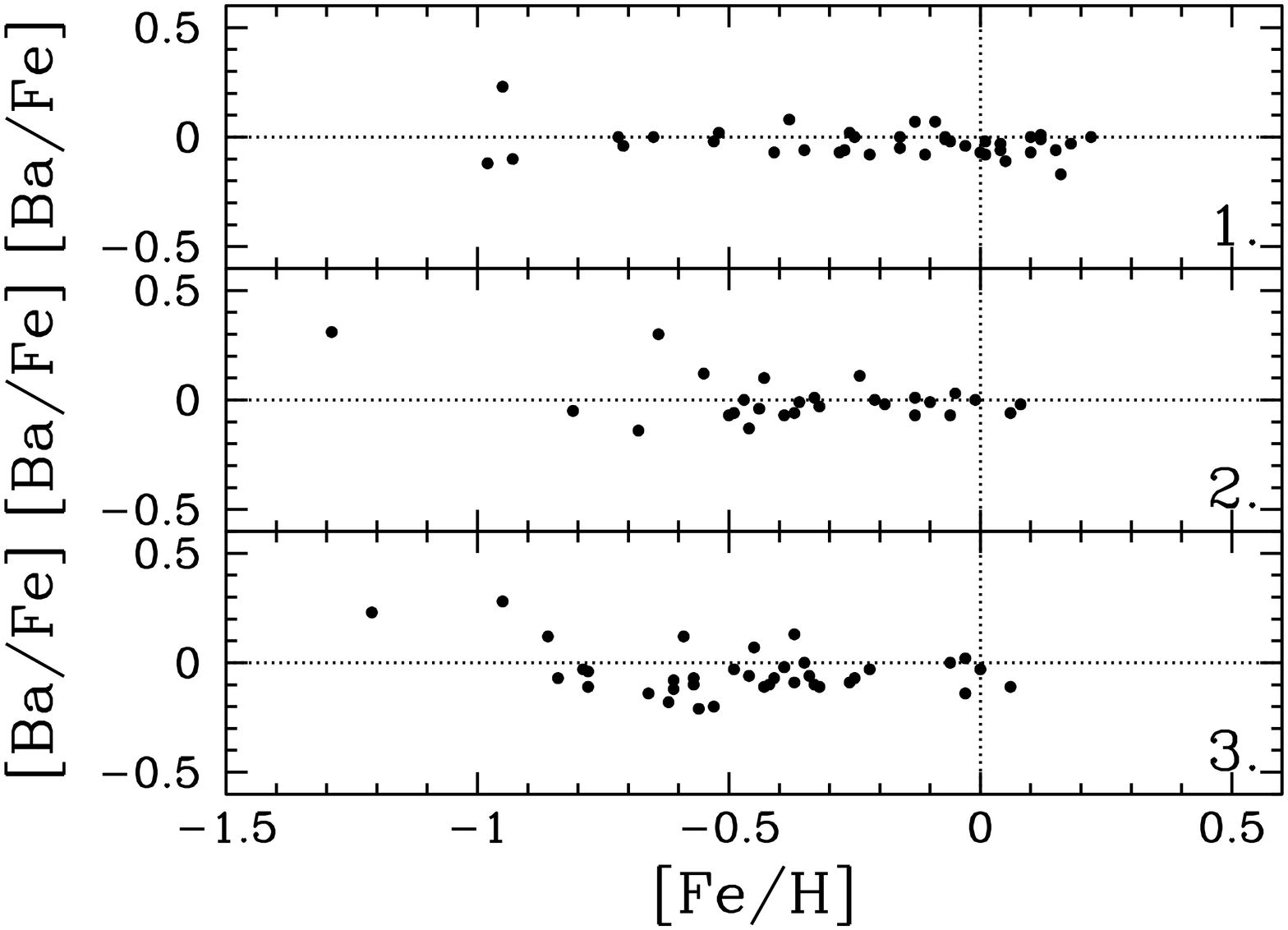}}
   \caption{[Ba/Fe] vs. [Fe/H] for the three samples defined in
     \fref{figze}. The numbering of the panels corresponds to the
     numbering of the boxes in \fref{figze}.  We do not include stars
     with only one Ba line measured or with large errors (we only
     select stars with $\sigma_{\rm line-to-line}/n_{\rm lines}
     <0.075$).}
   \label{bathree}
   \end{figure}

   \begin{figure}
   \centering
   \resizebox{\hsize}{!}{\includegraphics[angle=-90]{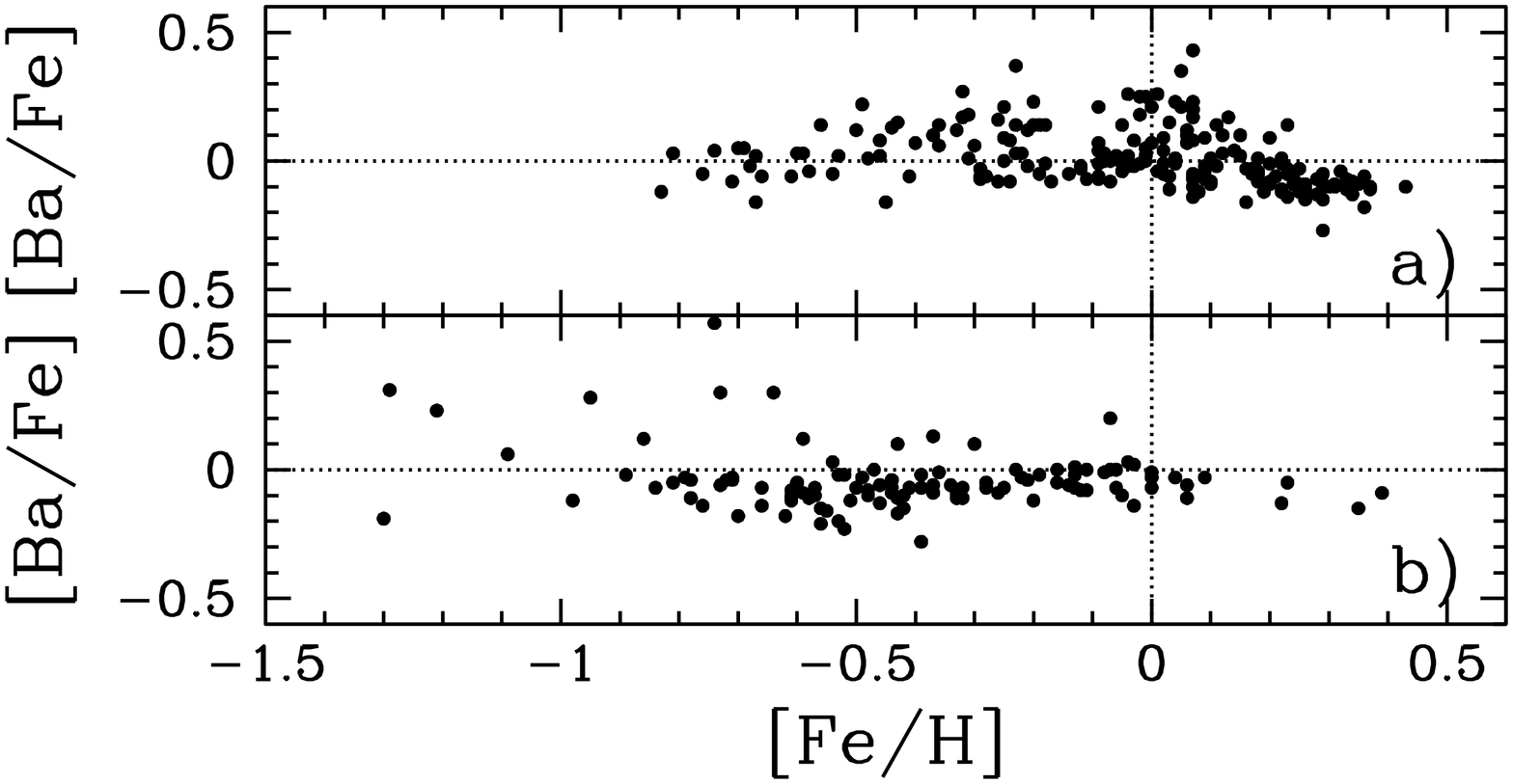}}
   \caption{[Ba/Fe] vs. [Fe/H] for {\bf a.} stars that are ten times
     more likely to be thin than thick disc and {\bf b.} stars that
     are ten times more likely to be thick than thin disc stars.  The
     separation is based on probabilities calculated according to the
     prescriptions in \citeasnoun{2003A&A...410..527B}. We do not
     include stars with only one Ba line measured or with (we only
     select stars with $\sigma_{\rm line-to-line}/n_{\rm lines}
     <0.075$). }
   \label{batt}
   \end{figure}

   Other elements also show tight abundance trends, further
   strengthening the suggestion that indeed (all or part of) the thick
   disc has formed from well-mixed gas. \Fref{bathree} show the
   [Ba/Fe] vs [Fe/H] trends for the three boxes defined in
   \fref{figze}. Again we see {\em very} tight abundance trends. A
   comparison with the trends for the thin and thick discs (same
   definition as used to create \fref{figtt} and \ref{batt}) shows
   that the trends in the three boxes mainly resemble those of the
   thick disc. The stars in Box 3 might have a tendency to show lower
   [Ba/Fe] than the stars in Box 1 and 2. This is certainly the case
   for [Fe/H] around --0.55. However, some care is needed in the
   interpretation of \fref{batt}a. In \citeasnoun{2007ApJ...663L..13B}
   we showed that at least some part of the apparent large scatter for
   the thin disc sample is due to what could be called an
   age-effect. It is the youngest stars that have the largest [Ba/Fe],
   in fact, [Ba/Fe] appears to be, for the last few billion years a
   monotonously increasing function of age. The older thin disc show a
   tight flat trend. Hence, the fact that the trend for Box 2
   resembles that for the thick disc more than that of the thin disc
   is not necessarily in contradiction with the observations of the
   thin disc.

\subsubsection{Discussion of Box 1 and 3}

It is generally understood that the heating mechanisms that work in
the plane, and thus cause higher eccentricity for the orbits, are
different from those that heat the disc vertically. The difference
manifest itself for example in that there is considerable structure in
the $U_{\rm LSR}$ and $V_{\rm LSR}$ plane but the $W_{\rm LSR}$
velocity shows a smooth distribution \cite{2007A&A...475..519H}. As
summarized in \citeasnoun{2007A&A...475..519H} the heating mechanisms
responsible for the vertical heating is observed to be more efficient
than what simulations of heating from within the disc can produce
(e.g. heating from molecular clouds and black holes).

The effect of a merger depends not only on the size of the in-falling
galaxy but also on its orbit
\cite{2008arXiv0806.2861H,2008arXiv0803.2714R}. \citeasnoun{2008arXiv0803.2714R}
explored a range of parameters and found that it is possible to create
stellar thick discs that are qualitatively similar to that seen in the
solar neighbourhood in the Milky Way.

We have investigated $Z_{\rm max}$ as a function of age for the stars
in Box 1 -- 3 as well as the thin and the thick disc stars.  There
does not appear to be any particular trend such that older stars on
average reach higher heights above the galactic plane. In the thick
disc sample there might be a slight tendency for this but over all age
and $Z_{\rm max}$ do not show any obvious correlations. If the stars
that now reach high above the galactic plane had acquired their high
$W_{\rm LSR}$ velocities over time (i.e. through heating from within
the disc itself) it would be natural to expect a correlation such that
stars that reach high above the plane are also the oldest, i.e. the
vertical velocity dispersion increases with age
\cite{2002ARA&A..40..487F}.  We do not appear to see that as
manifested in the distribution $Z_{\rm max}$ as a function of
age. Hence, a preliminary conclusion is that the major reason for the
observed $W_{\rm LSR}$ is not heating over time within the disc. This
is in agreement with dynamical studies
e.g. \citeasnoun{2002MNRAS.337..731H}. On the other hand, if the
observed velocities are the result of a merger event then that must
have happened quite late as also stars with ages of about 5 Gyr are
equally well represented at the high latitudes as older stars.  Hence
it appears natural to invoke an ``external'' heating mechanism such as
a major or minor merger to produce the observed velocity dispersion in
the vertical direction.

The very tight abundance trends observed for Box 1 and 3, and
especially for Box 3, in combination with the discussion above thus,
tentatively, leads to the conclusion that what we see in these two
Boxes is a stellar population that has been heated by a merger
event. Box 1 could be expected to be made up of mixture of stars.  It
should then contain stars that has never been heated in the vertical
direction but have suffered in-plane interactions that have put them
on more eccentric orbits.  Box 1 should then also contain stars that
in fact belong to Box 3 (i.e. heated stars) and either providing the
low velocity tail of that box or are stars that have lost their high
$W_{\rm LSR}$ thanks to encounters. The event that heated this
population should have happened a few billion years ago (keeping in
mind that our age determinations {\em de facto} are of a relative not
absolute nature).

\subsubsection{The stars in Box 2} In Box 2 we find that stars
that are on circular orbits (i.e. $e$ is low) but that
have high $W_{\rm LSR}$ show the same type of abundance pattern as
found for stars on solar type orbits (see \fref{figthree} and
\ref{figtt}). This directly implies that indeed these high $Z_{\rm
  max}$ stars originate from the solar neighbourhood. It would seem
natural to assume that these stars should be amongst the oldest stars
with solar type $U_{\rm LSR}$ and $V_{\rm LSR}$ velocities. This is,
however, not the case. The stars show a range of ages, just like the
stars with high likelihood to be thin disc stars. 

However, the stars in Box 2 show another feature -- there are no stars
with [Fe/H] (or indeed [Ti/H]) above solar amongst them, whilst our
thin disc sample has a large number of such stars (compare
e.g. \fref{batt}). Thus it appears that if the stars occupying Box 2
originally are stars from the solar neighbourhood with solar type
orbits that has since been vertically heated this heating took place
before the full chemical enrichment as seen in the thin disc took
place. If the heating in the form of a merger is needed in order to
explain the orbits of the stars in Box 2 this observations might thus
indicate that this heating event took place some time ago. However, an
answer to when would need detailed modeling of the chemical and
dynamical evolution. Also, the number of stars in Box 2 is not very
large and a larger sample should be gathered before firm conclusions
can be made.

\subsection{Orbit switching?}

   \begin{figure}
   \centering
   \resizebox{\hsize}{!}{\includegraphics[angle=-90]{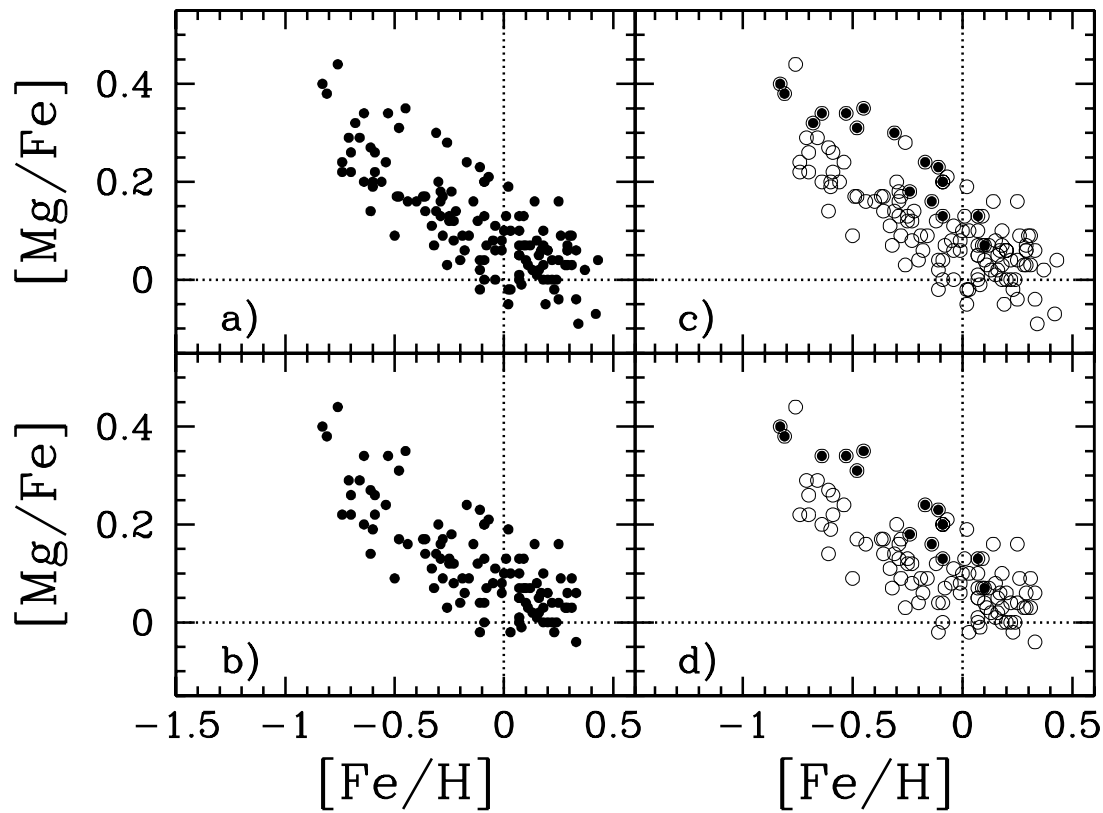}}
   \caption{Sign of migration? {\bf a.} Stars that are ten times more
     likely to be thin than thick disc star and which have relative
     errors in their age determinations less than 30\%. {\bf b.}
     Stars with $0<e<0.15$ and $Z_{\rm max}<0.5$ kpc and relative
     errors in their age determinations less than 30\%. {\bf c.} same
     as a. but with stars with age larger than 9 Gyr marked with
     $\bullet$. {\bf d.} same as b. but with stars with age larger
     than 9 Gyr marked with $\bullet$.}
   \label{figmig}
   \end{figure}

In \fref{figmig}a we show [Mg/Fe] vs [Fe/H] for the stars that are ten
times more likely to be thin than thick disc stars. Although the
general trend is of a single downward slope we find that there are a
set of stars that appear to show essentially the same abundance trend
but slightly ``elevated''. In \fref{figmig}c we have marked the stars 
that are older than 9 Gyr. Clearly, the stars that are extra-enhanced in
[Mg/Fe] are the oldest stars. We have explored if other cuts in age 
would produce other distinct trends but have found none. 

What is the origin of these stars? Could these stars have migrated
from the inner disc?  Could they be the remnants of an engulfed dwarf
galaxy?

\citeasnoun{2002MNRAS.336..785S} showed that it is possible for stars
to migrate significant distances radially in the stellar disc. This
means that stars that formed from promptly enriched gas in the inner
disc can have migrated out to the solar radius. A recent simulation,
investigating the truncation of stellar discs seen in other galaxies
show the same effect \cite{2008arXiv0807.1942R}. The migration of
stars would lead to a weakening of any age-metallicity relation in the
stellar disc and would, potentially, also confuse the elemental
abundance trends as stars that are the result of star formation
histories that have taken place at different positions in the Milky
Way disc end up on orbits that are too similar to be distinguished
kinematically. If the observed old, metal-rich stars with thin disc
but elevated [Mg/Fe] and thin disc kinematics actually originate in
the inner disc then, if migration is a constantly on-going process,
why are we seeing only one such trend? Does this imply that the orbit
switching happens only sometimes and for short periods of time?

In spite of the intriguing possibilities discussed above it appears
prudent to, for now, refrain from further speculations and conclude
that the exact nature of these few stars remains unclear and requires
further investigations.

\subsection{The metal-rich thick disc}

In \citeasnoun{2007ApJ...663L..13B} we asked the question: ``How
metal-rich can the thick disc be?'' For this very purpose we had
selected a sample of metal-rich (as judged from metallicities based on
photometry) stars with typical thick disc kinematics and obtained
high-resolution (R$\simeq$65,000), high signal-to-noise
(S/N$\simeq$250) spectra for F and G dwarf stars sampling the relevant
parameter space (for details see \citeasnoun{2007ApJ...663L..13B} and
Bensby et al. in prep.). The abundance analysis showed that indeed the
stars are metal-rich. Using a combination of kinematic and abundance
criteria we showed that stars with typical thick disc kinematics and
typical thick disc abundance patterns reached solar
metallicity. Moreover, these stars are also old. Older than stars with
the same [Fe/H] but with typical thin disc kinematics and typical thin
disc abundance patterns.

In \fref{figone}b the Toomre diagram for all stars with [Fe/H]$>$0 is
shown.  A large fraction of the stars are clustered together with
kinematics akin to that of the sun (the over-density centred at
$V_{\rm LSR}$$\simeq$0 km\,s$^{-1}$). But there are quite a few stars
at $V_{\rm LSR}$$<$--50 km\,s$^{-1}$ (remember that the asymmetric
drift of the thick disc is about --50 km\,s$^{-1}$), indeed there are
stars all the way down to about --100 km\,s$^{-1}$ and stars with high
$U_{\rm LSR}$ and $W_{\rm LSR}$ velocities. Twelve of these stars
would be selected as ten times more likely to be thick disc stars than
thin disc stars using the scheme in
\citeasnoun{2003A&A...410..527B}. These stars have $V_{\rm tot}$
around 100 km\,s$^{-1}$ or larger. 

This simple, kinematic exploration of stars with super-solar [Fe/H]
shows that the solar neighbourhood contains metal-rich, high velocity
stars that very likely are associated with the thick disc. Proper dynamical
modeling of such stellar populations are necessary in order to fully
understand the origin and nature of the thick disc. 

\section{Moving groups and stellar streams}
\label{streams}

The stars in the solar neighbourhood are not smoothly distributed in
the $U_{\rm LSR}$ -- $V_{\rm LSR}$ plane. Indeed, significant
sub-structure is present and several so called moving groups and
stellar streams have been identified through detailed analysis of the
{\it Hipparcos} data
\cite{1998AJ....115.2384D,2005A&A...430..165F,2006A&A...449..533A}.
However, the concept of moving groups is older than these
investigations and was originally introduced by O. Eggen, although his
definitions of the groups have been largely superseded by the work
using the {\it Hipparcos} parallaxes.

The origin of these moving groups and stellar streams remain
unclear. Are they evaporating stellar clusters, accreted dwarf galaxies
or the result of secular evolution in the Milky Way disc? Studies of
the elemental abundances and ages for the stars in the groups may hold
the answer.

\paragraph{The Hercules and Arcturus stellar streams}
The presence of streams in the Milky Way disc, like the Hercules and
Arcturus streams, has lead to the speculation that they may be the
remnants of a minor merger, a dwarf galaxy that has been engulfed by
the Milky Way disc and now is only visible as perturbations in the
velocity field. Models of minor mergers show that the stars may indeed
end up on orbits typical for the stars in these streams 
\cite{williamsphd2008}.
However, detailed dynamical modeling have instead
attributed these streams to dynamical process in the Milky Way disc,
probably partly driven by the presence of a bar in the central parts
of the Galaxy \cite{1998AJ....115.2384D}.

The Hercules stream was found to contribute about 6\% of all stars in
the solar neighbourhood \cite{2005A&A...430..165F}.  The stream has a
net drift relative to the local standard of rest of about $\sim$40 km
s$^{-1}$ directed radially away from the galactic centre. It also has
a assymetric drift similar to that of the thick disc.  The Hercules
stream is included in our sample and can be seen as an increase in the
stellar density at $V_{\rm LSR} < -60$km s$^{-1}$ in \fref{figone}.

Two recent studies provide detailed abundance analysis of stars in
these streams 
\cite{2007ApJ...655L..89B,williamsphd2008}.
In both cases neither stream show a
distinct abundance or age pattern. Instead, the stars associated with
the stream merges into the general trends found for the disc stars in
general. In fact \citeasnoun{2007ApJ...655L..89B} find that the stars
in the Hercules stream falls essentially on the two age-metallicity
trends traced by their thin and thick disc samples, respectively. 

The results from these studies thus imply that the two streams
indeed are the results of dynamical evolution within the Milky Way
disc itself rather than the merger of a smaller dwarf galaxy with the
Milky Way.

\paragraph{The HR1614 moving group}
The HR1614 moving group 
was originally defined by Eggen, see for example
\citeasnoun{1992AJ....104.1906E} and references therein. Recently,
\citeasnoun{2007AJ....133..694D} obtained high resolution, high S/N
spectra for members of the HR1614 moving group. Their abundance analysis
revealed that indeed the stars within this group show extremely
homogeneous elemental abundances. In fact, their sample contained a few
field stars which were easily identifiable thanks to their discrepant
abundances.

Hence, so far it is only the HR1614 moving group that has shown itself
to be an homogeneous, single age, single abundance stellar
population. The other groups appear to be the kinematic gathering of
stars that have suffered secular evolution within the disc and are not
the result of dissolving stellar clusters or captured satellite
galaxies. These results are in accordance with a new study by 
\cite{antoja2008}. They find that indeed the stars in most of
the easily identifiable structures in the $U_{\rm LSR}$ -- $V_{\rm
  LSR}$ plane do not share a common age or metallicity, but instead
show significant spreads in these parameters.

\section{Summary}

We have presented data for a new stellar sample, comprising some 550 F
and G dwarf stars. These stars span a large range of orbital
parameters and we have been able to explore the Milky Way stellar
disc, as manifested in the solar neighbourhood, in some detail.
In particular we find the following

\begin{itemize}

\item We find remarkably tight abundance trends for stars that reach
  high above the galactic plane and have eccentric orbits show
    
\item A tentative interpretation of these abundance trends is that the
  gas that these stars formed out of must have been well mixed on
  large scales (on the order of kpc)

\item We confirm previously found abundance trends for statistically
  selected samples of thin and thick disc stars

\end{itemize}

Additionally, we discuss the presence of a significant number of
high-velocity stars with super-solar metallicities as well as a
possible detection of stars that are tentative candidates for orbit
switching, i.e. stars that were born on a circular orbit further in
towards the galactic centre but now have been moved out to new, still
circular, orbits in the solar neighbourhood.

Recent studies employing high resolution abundances analysis for so
called moving groups and stellar streams are starting to fill an
important gap in our knowledge about the origin of the ``lumpy'' parts
of the velocity distributions in the Milky Way disc. The first results
show that these groups and streams are an heterogeneous set of
objects. The HR\,1614 moving group show very coherent elemental
abundances for its stars. This group of stars could definitely be
found by using so called chemical tagging
\cite{2002ARA&A..40..487F}. The Arcturus and Hercules streams on the
other hand seem to simply be the result of dynamical processes within
the stellar disc, perhaps powered by the bar. It remains important to
explore additional groups and streams. These studies should be done in
a differential manner to the general disc population in order to
better understand their origin.

Extra-galactic observation provide more and more evidence for ordered
old and metal-rich stellar and gaseous discs in large galaxies at high
redshifts. Locally, in the Milky Way, we find that stars on ``hot'',
thick disc like orbits show extremely tight abundance trends
indicating that they were formed out of gas that was well-mixed over
large scales. Given that these stars, on average, also are old we are
potentially seeing the local counterpart to those high redshift discs.

\bigskip

{\em Finally, I would like to thank all the organizers of this
  meeting. You had truly managed to create a meeting in the spirit of
  Bengt Gustafsson's multi-faceted achievements within and outside of
  academe. Many thanks. 

Last I would like to thank Bengt -- without you, ofcourse, we would
not have had this inspiring and interesting meeting. A special
thanks to you and Sigbritt for treating us to such a wonderful concert! }

\ack 
SF is a Royal Swedish Academy of Sciences Research Fellow
supported by a grant from the Knut and Alice Wallenberg Foundation.
This work was supported
in part by the U.S. National Science Foundation, grant AST-0448900.

\section*{References}
\bibliographystyle{jphysicsB}
\bibliography{feltzing2}

\end{document}